\documentclass[aps,amssymb,letterpaper,amsmath,twocolumn]{revtex4}
\usepackage{cancel}
\usepackage{ulem}
\begin{document}
\title{Density Matrix Renormalization Group Lagrangians}
\author{Garnet Kin-Lic Chan}
\email[]{E-mail:gc238@cornell.edu}
\affiliation{Department of Chemistry and Chemical Biology, Cornell
  University, Ithaca, New York 14853, USA}
\date{\today}
\begin{abstract}
We introduce a Lagrangian formulation of the Density Matrix
Renormalization Group (DMRG). We present Lagrangians  which when
minimised 
yield the optimal  DMRG wavefunction in a variational sense, both within the
general matrix product ansatz, as well as within the  canonical form of the
matrix product that is constructed within the DMRG
sweep algorithm. Some of the results obtained 
are similar to elementary expressions in Hartree-Fock theory,
and we draw attention to such analogies. The Lagrangians introduced
here will be useful in developing theories of analytic response and
derivatives in the DMRG.
\end{abstract}

\maketitle
\newcommand{\boldpsi}{\boldsymbol{\psi}}

\section{Introduction}



The density matrix renormalisation group (DMRG) of White
\cite{White1992, White1993, Schollwock2005} is a recent addition to the
methods of quantum chemistry \cite{White1999, Mitrushenkov2001,
  Chan2002, Legeza2003dyn, Moritz2005orb, Zgid2008, Chan-bookchapter}. Unlike many other correlation methods the DMRG
is not based on excitations from  a Hartree-Fock reference 
but rather on a new kind of highly flexible reference function. In
quantum chemistry, it has led to advances in the treatment of   strongly
interacting (i.e. multi-reference)  problems. For 
molecules that are large in one  spatial dimension, 
the cost of the DMRG is only quadratic in
the number of localised orbitals  and it is   therefore  a quadratic-scaling multi-reference
method for such systems \cite{Hachmann2006}. We  have applied a quadratic-scaling DMRG
algorithm to study   conjugated polymers \cite{Hachmann2007, Dorando2007}, light-harvesting pigments \cite{Ghosh2008}, and the metal-insulator
transition in hydrogen chains \cite{Hachmann2006}  with full treatment of multi-reference
correlations in as large as 100 electron, 100 orbital complete active spaces.




Early formulations of the DMRG  primarily used  the  language of the
numerical renormalisation group that  reflects the  history of its
development. Such language is very different from the usual language
of quantum chemistry. However, as is now understood, the DMRG
algorithm  simply minimises the energy of  a wavefunction ansatz  known as the matrix product
state \cite{Ostlund1995, Rommer1997, Schollwock2005, Chan-bookchapter}. This ansatz has a very different structure from most quantum
chemical wavefunctions and the unique strengths and weaknesses of the
DMRG method can be understood from this point of view \cite{Chan-bookchapter}.

In a prior publication \cite{Chan-bookchapter} we have presented an introduction to the DMRG
from the wavefunction perspective. In the current work, we continue along this line of presentation and
describe  simple reformulations of the DMRG that connect the
method with well-known \textit{Lagrangian} techniques in quantum
chemistry \cite{Helgaker1989, Helgaker1989nsp, Koch1990}. All our results are of a very  elementary nature, but we feel there is
sometimes value to writing  out such things
explicitly.  In the past, Lagrangian formulations have provided a unified and
systematic language by which to derive many results in the area of
perturbation theory and analytic derivatives and response \cite{Helgaker1989,Helgaker1989nsp,Koch1990}. We
believe that the Lagrangian formulation of the DMRG presented here
will be useful in a similar way. 

We start in section \ref{sec:dmrg_wave} by recalling the  matrix product structure of the
DMRG wavefunction. In section \ref{sec:dmrg_lagrange} we write
down a simple Lagrangian for the DMRG wavefunction and  the
corresponding  stationary equations. These resemble the
the Fock orbital equations of Hartree-Fock theory and we discuss this
similarity. In DMRG calculations that are based on the traditional
sweep algorithm, one implicitly uses not the
most general form of the DMRG ansatz, which contains some redundancy, 
but rather a special \textit{canonical} form \cite{Chan-bookchapter}. The canonical form of
the DMRG wavefunction is reviewed in the first part of section
\ref{sec:dmrg_canonical} while in the second part we introduce the 
additional constraints  that
have to be applied to Lagrangian to ensure that the minimising
wavefunction  is of canonical form. We finish  by demonstrating explicitly that minimising
 the canonical DMRG Lagrangian  yields  exactly the same solution
conditions on the DMRG wavefunction as the original sweep algorithm of
the DMRG.

\section{The DMRG wavefunction }

\label{sec:dmrg_wave}
 
Recall the full configuration interaction expansion of the wavefunction
in Fock space. In terms of Slater determinants written in the
occupation number representation $|n_1 \ldots n_k\rangle$, where $n_i$
is the occupation of orbital $i$ taking values $0, 1_\alpha, 1_\beta,
1_{\alpha}1_\beta$, this is
\begin{align}
|\Psi\rangle &=  \sum_{n_1 n_2 n_3 \ldots n_k}\Psi^{n_1n_2n_3 \ldots n_k}
|n_1 n_2 n_3 \ldots n_k\rangle \label{eq:fci}, \\
 \sum_{i} n_i & = N.
\end{align}


In the DMRG ansatz, the
  expansion coefficient $\Psi^{n_1 n_2 n_3 \ldots n_k}$ is represented
  by a contracted product of the tensors, where each tensor is
  associated with the Fock space of a given orbital. In the context of
  the DMRG it is more usual to refer to orbitals as sites, and thus we
  refer to the tensors as site-functions. Thus we write
\begin{align}
\Psi^{n_1 n_2 n_3 \ldots n_k}= \psi^{n_1}_{i_1} \psi^{n_2 i_1}_{i_2}
\psi^{n_3 i_2}_{i_3} \ldots
\ldots \psi^{n_k}_{i_{k-1}} \label{eq:mps}
\end{align}
where we have used the Einstein summation convention that we will
employ throughout this work (i.e. repeated upper and lower indices are
contracted). 
The number of coefficients in each site-function (save for the first
and last) is $4 M^2$, where $M$ is the dimension of each  $i$
index. (It is conventional to take the dimension of each $i$ index to
be the same). 




From eqn. (\ref{eq:mps}) we see that the DMRG ansatz has a contracted
matrix product structure. For this reason it is known as a
matrix product state \cite{Ostlund1995, Rommer1997, Schollwock2005,
  Chan-bookchapter}. The matrix product maybe used to reconstruct the
Slater determinant expansion of the DMRG wavefunction (see e.g. \cite{Moritz2007}).
The product nature is reminiscent of the orbital product ansatz in
Hartree-Fock theory. However, there are some important 
differences. Firstly, the number of site-functions is the size of the
basis $k$, rather than the number of electrons $N$. Thus the product
structure of the DMRG is expressed in the full Fock space, not in the
$N$-particle Hilbert space. Secondly, the ansatz is a contracted product rather than
a simple product. Thus correlations are introduced between the
orbital Fock spaces, by virtue of the contraction structure of  the
$i$ indices in the ansatz.

\section{The DMRG Lagrangian}

\label{sec:dmrg_lagrange}

We can  determine the best site-functions in the DMRG ansatz in a variational sense by minimising the energy subject to normalisation of the
wavefunction. The corresponding Lagrangian is
\begin{align}
\mathcal{L}[\Psi] = \langle \Psi|\hat{H} |\Psi\rangle  - E\langle
\Psi|\hat{1}|\Psi\rangle \label{eq:dmrg_lagrangian}
\end{align}
To evaluate the Lagrangian explicitly with the DMRG ansatz
we use a Fock  representation of the Hamiltonian. The matrix elements
of the Hamiltonian are written as
 \begin{equation}
 \langle n_1 n_2 \ldots n_k | \hat{H} | n_1^\prime n_2^\prime \ldots
 n_k^\prime \rangle = H^{n_1 \ldots n_k}_{n_1^\prime \ldots n_k^\prime}
 \end{equation}
In terms of the site-functions, the energy term $\langle
\Psi|\hat{H}|\Psi\rangle$ becomes
\begin{align}
\left(
\psi^{n_1}_{i_1} \psi^{n_2 i_1}_{i_2} \ldots
\psi^{n_k}_{i_{k-1}} 
\right)
H^{n_1 \ldots n_k}_{n_1^\prime \ldots n_k^\prime}
\left(
\psi_{n_1^\prime}^{i_1^\prime} \psi_{n_2^\prime
  i_1^\prime}^{i_2^\prime} \ldots \psi_{n_k^\prime}^{i_{k-1}^\prime}
\right)
\end{align}
while the normalisation term $\langle \Psi | \hat{1}|\Psi\rangle$ is 
\begin{align}
\left(\psi^{n_1}_{i_1} \psi^{n_2 i_1}_{i_2} \ldots \psi^{n_k}_{i_{k-1}}\right)
\left(\psi_{n_1}^{i_1} \psi_{n_2 i_1}^{i_2} \ldots \psi_{n_k}^{i_{k-1}}\right)
\end{align}
Note that each term in the Lagrangian is  quadratic in each of the site-functions. At the minimum, the derivative of the Lagrangian with
respect to the site functions vanishes. 
Then, the stationary equation satisfied by each site function is
\begin{align}
F^{n^\prime_p i^\prime_{p-1} i_p }_{i^\prime_p n_{p} i_{p-1} }
\psi^{n_p i_{p-1}}_{i_p} = E \psi^{n^\prime_p i^\prime_{p-1}}_{i^\prime_o}
\end{align}
with the matrix elements of the operator $F[p]$ defined as
\begin{align}
F^{n^\prime_p i^\prime_{p-1} i_p }_{i^\prime_p n_{p} i_{p-1} }& =  
\left(
\psi^{n_1}_{i_1}  \ldots \cancel{\psi^{n_p i_{p-1}}_{i_p}} \ldots
\psi^{n_k}_{i_{k-1}} 
\right) \times \nonumber \\ 
& H^{n_1 \ldots n_k}_{n_1^\prime \ldots n_k^\prime}
\left(
\psi_{n_1^\prime}^{i_1^\prime} \ldots \cancel{\psi_{n_p^\prime
  i_{p-1}^\prime}^{i_p^\prime}} \ldots \psi_{n_k^\prime}^{i_{k-1}^\prime}
\right) \\
&= \langle \Psi|\hat{H}|\Psi\rangle_{\cancel{p}} \label{eq:Fock}
\end{align}
where in the first line the struck-out symbols indicate that the corresponding site-functions are
omitted from the sum, and this is denoted also by the more compact notation in the
second line.

The stationary equations for the site-functions are analogous to the orbital
Fock equations of Hartree-Fock theory \cite{szabo1989mqc} as each site-function is an eigenfunction of an effective site ``Fock'' operator
$F[p]$. However, unlike in Hartree-Fock theory, the Fock operator is
different for each site, and all site-functions possess the
same eigenvalue $E$.  The site Fock operator $F[p]$ may be decomposed into local-site and
off-site terms. Assuming the usual form of the  electronic
Hamiltonian 
 \begin{equation}
 \hat{H} =  t^{ij} a^\dag_i a_j + v^{ijkl} a^\dag_i
 a^\dag_j a_k a_l
 \end{equation}
where for simplicity we are assuming summations over the spin-labels
of the orbitals, i.e. $t^{ij} a^\dag_i a_j = t^{i\sigma j\sigma^\prime} a^\dag_{i\sigma} a_{j\sigma}$.
We define the local-site contribution to $F[p]$ as
\begin{equation}
F[p]^{\text{(local)}}=\langle \Psi | t^{pp} a^\dag_p a_p + v^{pppp}
a^\dag_p a^\dag_p a_p a_p | \Psi \rangle_{\cancel{p}}
\end{equation}
and the off-site contributions as
\begin{align}
F[p]^{ \text{(off-site)}}&=\langle \Psi | t^{ij} a^\dag_i a_{j\text{(not $i=j=p$)}}+ \nonumber \\ &v^{ijkl} a^\dag_i
a^\dag_j a_k a_{l \text{(not $i=j=k=l=p$)}} | \Psi \rangle_{\cancel{p}}
\end{align}
This division is analogous to the division of the Fock operator
into one-electron and two-electron Coulomb-exchange terms. In particular, the
off-site contributions represent the contributions of the average
``field'' of all the sites to the local Fock operator at site $p$.

\section{The Canonical DMRG Lagrangian}

\label{sec:dmrg_canonical}

\subsection{Canonical form of the DMRG wavefunction} 
The DMRG wavefunction as written in (\ref{eq:mps})
possesses many redundant degrees of freedom. For example, given an
arbitrary invertible matrix $T$, we can obtain multiple equivalent
matrix product approximations for the
wavefunction tensor $\Psi^{n_1 \ldots n_k}$ by inserting $T, T^{-1}$
in between two site functions, e.g.
\begin{align}
\Psi^{n_1 \ldots n_k}& = \psi^{n_1}_{i_1} \ldots \psi^{n_p  i_{p-1}}_{i_p}
\psi^{n_{p+1} i_p}_{i_{p+1}} \ldots \psi^{n_k}_{i_{k-1}} \nonumber \\
&=\psi^{n_1}_{i_1} \ldots \psi^{n_p  i_{p-1}}_{i_p} \left(T^{i_p}_{ i_p^\prime}
{T^{-1}}^{i_{p^\prime}}_{i_p^{\prime\prime}}\right)\psi^{n_{p+1}
i_p^{\prime\prime}}_{i_{p+1}} \ldots \psi^{n_k}_{i_{k-1}} \label{eq:wf_equivalence}
\end{align}
Thus minimisation of the DMRG Lagrangian (\ref{eq:dmrg_lagrangian})
does not define the site-functions uniquely, but  only up to 
pairs of transformations \cite{Chan-bookchapter, Ghosh2008}.


The original sweep
algorithm used to optimise the DMRG wavefunction does, however, define a particular
choice of site-functions at convergence. These site-functions
are canonical in ways which resemble the properties of canonical orbitals in Hartree-Fock theory. 
In Hartree-Fock theory, the canonical orbitals
diagonalise the Lagrange multipliers associated with orbital
orthonormality. As we show in the next section 
the canonical site functions obtained from 
the DMRG sweep algorithm   diagonalise a matrix of
Lagrange multipliers associated with orthogonality constraints.


Let us first recall how the sweep algorithm leads to a canonical form
of the DMRG wavefunction and site-functions. We will then extract the solution
conditions satisfied by the canonical site-functions at the
convergence of the sweep algorithm. 
We will assume here some familiarity with the DMRG sweep algorithm and
we refer readers to our earlier work and review for a complementary
discussion \cite{Chan2002, Chan-bookchapter}. (In particular, here we
will focus on the ``one-site'' variant of the DMRG algorithm
\cite{Chan2002, White2005, Zgid2008onesite}).

In the sweep algorithm the site-functions are
seen as transformation matrices which define sets of renormalised
bases. For example, at block configuration 
 $\fbox{$\bullet_1 \ldots \bullet_{{p-1}}$} \bullet_p
   \ \fbox{$\bullet_{p+1} \ldots \bullet_k$}$, the first
   $p-1$ site-functions define many-body ``left'' basis functions recursively
   through
\begin{align}
|l_{p-1}\rangle& = \sum_{n_{p-1} l_{p-2}} L^{n_{p-1} l_{p-2}}_{l_{p-1}} |n_{p-1} l_{p-2}\rangle
 \nonumber \\
& = \sum_{n_1 \ldots n_{p-1}} L^{n_1}_{l_1} L^{n_2 l_1}_{l_2} \ldots L^{n_{p-1}
  l_{p-2}}_{l_{p-1}} |n_1 \ldots n_{p-1}\rangle
\end{align}
while site-functions $p+1 \ldots k$ define many-body ``right'' basis
functions recursively through
\begin{align}
|r_p\rangle& = \sum_{n_{p+1} r_{p+1}} R^{n_{p+1} r_{p+1}}_{r_p} |n_{p+1} r_{p+1}\rangle
 \nonumber \\
& = \sum_{n_{p+1} \ldots n_{k}} R^{n_{p+1} r_{p+1}}_{r_p} \ldots
R^{n_{k-1} r_{k-1}}_{r_{k-2}} R^{n_k}_{r_{k-1}} |n_{p+1} \ldots n_k \rangle
\end{align} 
The transformation matrices are orthogonal in the sense that $\langle l_{p-1}|
l^\prime_{p-1} \rangle = \delta^{l_{p-1}}_{l_{p-1}^\prime}$ and
similarly for the right basis functions; this implies 
\begin{align}
L^{l_q}_{n_q l_{q-1}} L^{n_q l_{q-1}}_{l_q^\prime}& =
\delta^{l_q}_{l_q^\prime} \nonumber \\
R^{r_{q-1}}_{n_q r_q} R^{n_q r_{q}}_{r_{q-1}^\prime}& =
\delta^{r_{q-1}}_{r^\prime_{q-1}} \label{eq:ortho}
\end{align}
Using these definitions of the left and right bases $\{
l_{p-1} \}$, $\{ r_p\}$ as well as the basis of site $p$, $\{ n_p \}$
the total wavefunction at the block configuration  $\fbox{$\bullet_1 \ldots \bullet_{{p-1}}$} \bullet_p
   \ \fbox{$\bullet_{p+1} \ldots \bullet_k$}$ is expanded  as
\begin{align}
|\Psi\rangle& = \sum_{l_{p-1} n_p r_p} C^{l_{p-1} n_p r_{p}}   |l_{p-1} n_p r_{p}
 \rangle  \nonumber \\
&=\sum_{n_1 \ldots n_k} L^{n_1}_ {l_1} \ldots L^{n_{p-1} l_{p-2}}_{l_{p-1}} \times \nonumber \\
& C^{l_{p-1} n_p  r_p} R^{n_{p+1} r_{p+1}}_{r_p} \ldots R^{n_k}_{
  r_{k-1}}  |n_1 \ldots n_k\rangle \label{eq:renorm_wf} 
\end{align}
We see that the form of the wavefunction constructed in the sweep algorithm has a matrix product structure as in
eqn. (\ref{eq:mps}) but has additional orthogonality constraints on the
site-functions (\ref{eq:ortho}). Also, this wavefunction provides 
a special meaning to the $p$th site-function $C^{l_{p-1} n_p r_{p}}$, which appears as the set
of expansion coefficients associated with the renormalised product basis $\{
l_{p-1} n_p r_{p} \}$. We refer to the 
DMRG wavefunction constructed in the form (\ref{eq:renorm_wf}) as the
site $p$ \textit{canonical form} \cite{Chan-bookchapter, Ghosh2008}.



We now recall how the the  site-functions appearing in the site
$p$ canonical form of the wavefunction (\ref{eq:renorm_wf}) are
determined in the  sweep algorithm. The coefficients $C^{l_{p-1}
  n_p r_{p}}$ are obtained by solving the Schr\"odinger
equation projected into the product basis $\{ l_{p-1} n_p r_{p}\}$
\begin{align}
\langle l_{p-1}^\prime n_p^\prime r_{p}^\prime |\hat{H} |l_{p-1} n_p r_{p}
 \rangle C^{l_{p-1} n_p r_{p}} &= E C^{l_{p-1}^\prime n_p^\prime
   r_{p}^\prime} \label{eq:renorm_se}
\end{align}
These coefficients  determine corresponding
$L$, $R$ site-functions at the same site $p$ (i.e. $L^{n_p l_{p-1}}_{l_p}$ and $R^{n_p
  r_{p}}_{r_{p-1}}$) as eigenvectors of 
appropriate reduced density matrices. For example, $L^{n_p
  l_{p-1}}_{l_p}$ is obtained from the eigenvectors of a
density matrix $\Gamma^{n_p l_{p-1}}_{n_p^\prime l_{p-1}^\prime}$ constructed by tracing over the right indices of the
wavefunction coefficients
\begin{align}
\Gamma^{n_p l_{p-1}}_{n_p^\prime l_{p-1}^\prime}& = C^{l_{p-1} n_p
  r_{p}} C_{l_{p-1}^\prime n_p^\prime r_{p}} \\
\Gamma^{n_p l_{p-1}}_{n_p^\prime l_{p-1}^\prime} L^{n_p^\prime
  l_{p-1}^\prime}_{l_p}& = w_{l_p} L^{n_p l_{p-1}}_{l_p} 
\label{eq:left_dm}
\end{align}
while $R^{n_{p} r_{p+1}}_{r_p}$ is obtained from the eigenvectors of a
density matrix $\Gamma^{n_p r_{p}}_{n_p^\prime r_{p}^\prime}$ obtained by tracing over the left indices
\begin{align}
\Gamma^{n_p r_{p}}_{n_p^\prime r_{p}^\prime} = C^{l_{p-1} n_p
  r_{p}} C_{l_{p-1} n_p^\prime r_{p}^\prime}\\
\Gamma^{n_p r_{p}}_{n_p^\prime r_{p}^\prime} R^{n_{p}^\prime
  r_{p}^\prime}_{r_{p-1}} = w_{ r_p-1}  R^{n_{p}
  r_{p}}_{r_{p-1}} \label{eq:right_dm}
\end{align}
The $L$ and $R$ site-functions at site $p$ do not
themselves appear in the site $p$ canonical form; rather we need
the  $L$ site functions at sites $1 \ldots p-1$ and the $R$ site
functions at sites $p+1 \ldots k$. But these can be obtained by
solving the effective Schr\"odinger equation (\ref{eq:renorm_se}) 
at other  block configurations in the sweep. Sweeping
through  block configurations  $\fbox{$\bullet_1 \ldots \bullet_{{p-1}}$} \bullet_p
   \ \fbox{$\bullet_{p+1} \ldots \bullet_k$}$ for $p=1 \ldots k$,
and solving for the  wavefunction coefficients $C$ at each block
configuration, we can obtain all the $L$ and $R$ site
functions appearing in the site $p$ canonical form
(\ref{eq:renorm_wf}) \cite{Chan2002, Chan-bookchapter}.

Note that any wavefunction written in the canonical form of
one site (say $p$) can always be written exactly in the canonical form of
another site (say $q$). In this sense, canonical forms at different
sites are simply different representations of the same wavefunction \cite{Chan-bookchapter}. More precisely, given  
$C^{l_{q-1} n_q r_q}, L^{n_q l_{q-1}}_{l_q}$ at site $q$, we can always find
$C^{l_{p-1} n_p {r_p}}, R^{n_p r_{p-1}}_{r_{p}}$ at site $p>q$ such that
\begin{align}
L^{n_1}_{l_1} \ldots L^{n_q l_{q-1}}_{l_q} \ldots
C^{l_{p-1} n_p r_p} \ldots R^{n_k}_{ r_{k-1}} \nonumber \\
=
L^{n_1}_{l_1} \ldots C^{l_{q-1} n_q r_q} \ldots
{R^{n_p r_{p-1}}_{r_p}} \ldots R^{n_k}_{ r_{k-1}} \label{eq:wave_trans}
\end{align}
In the sweep algorithm, the conversion between the canonical
forms of the DMRG wavefunction at neighbouring sites is known as the
wavefunction transformation \cite{White1996, Chan2002, Chan-bookchapter}, and it is commonly used to accelerate
the convergence of the sweeps.
At convergence, if $C^{
  l_{p-1} n_p r_p}$ solves the effective Schr\"odinger equation
(\ref{eq:renorm_se}) at site
$p$, then the corresponding $C^{l_{q-1} n_q r_q}$ determined through
the wavefunction transformation  solves
the effective Schr\"odinger equation (\ref{eq:renorm_se}) at site $q$. 

Let us now summarise the  solution conditions satisfied by the site-functions
appearing in the site $p$ canonical form (\ref{eq:renorm_wf}) 
at the convergence of the
DMRG sweep algorithm. 
\begin{enumerate}
\item For a specified site ($p$, say), the wavefunction coefficients
  $C^{l_{p-1} n_p r_{p+1}}$ satisfy the effective Schr\"odinger
    equation (\ref{eq:renorm_se}) and satisfy the normalisation condition
$C^{l_{p-1} n_p r_{p+1}} C_{l_{p-1} n_p r_{p+1}}=1$,
\item The $L$ and $R$ site-functions are each orthogonal in the sense
  of (\ref{eq:ortho}) and are related to the 
  $C$ site-functions (in the corresponding
  canonical forms) as eigenvectors of the correponding  density matrices
   (\ref{eq:left_dm}),(\ref{eq:right_dm}),
\item The $C$ site-functions appearing in all the canonical forms
  from site $1 \ldots k$ are   related through the wavefunction transformation (\ref{eq:wave_trans}).
\end{enumerate}


\subsection{Lagrangian formulation}

Let us now show how the above conditions 1.-3. satisfied by  the
canonical site-functions at the convergence of the   sweep
algorithm can be obtained by  minimising an appropriate canonical Lagrangian. 
We first note that $C$ is constrained to have unit norm while the $L$,
$R$ site-functions are orthogonal in the sense (\ref{eq:ortho}). Thus
we write a Lagrangian with these constraints
\begin{align}
\mathcal{L}[\Psi]  &=  \langle \Psi | \hat{H} | \Psi \rangle 
- E  \left(C^{l_{p-1} n_p r_p} C_{l_{p-1} n_p r_p} - 1\right)
\nonumber \\
&-\sum_{q<p}  
\mu_{l_q}^{l_q^\prime} \left( L^{l_q}_{n_q l_{q-1}} L^{n_q l_{q-1}}_{l_q^\prime} -
\delta^{l_q}_{l_q^\prime} \right) \nonumber \\
&- \sum_{q>p} \mu_{r_{q-1}}^{r^\prime_{q-1}}   \left(
R^{r_{q-1}}_{n_q r_q} R^{n_q r_{q}}_{r_{q-1}^\prime} -\delta^{r_{q-1}}_{r^\prime_{q-1}}
\right) \label{eq:ortho_lagrange}
\end{align}


At the minimum, derivatives of the Lagrangian
with respect to all $L$, $C$, $R$ site-functions must vanish. 
Differentiating with respect to the  coefficients $C^{l_{p-1}n_p r_p}$, we obtain an effective Fock
eigenvalue equation for $C$ similar to  eqn. (\ref{eq:Fock})
\begin{align}
F^{n_p l_{p-1} r_p }_{n_p^\prime l_{p-1}^\prime r_p^\prime}
C^{l_{p-1}^\prime n_p^\prime r_p^\prime} = E C^{l_{p-1} n_p r_p} \label{eq:canonical_fock}
\end{align}
Comparing this with the effective Schr\"odinger
equation (\ref{eq:renorm_se}) we see that $F^{n_p l_{p-1} r_p
}_{n_p^\prime l_{p-1}^\prime r_p^\prime} = \langle l_{p-1}^\prime
n_p^\prime r_{p}^\prime |\hat{H} |l_{p-1} n_p r_{p}\rangle$, and thus
(\ref{eq:canonical_fock}) is simply the same as solution condition 1.
from the sweep algorithm.


Next we consider minimising $\mathcal{L}$ with respect to the left and right
site-functions. In each case there are two  non-vanishing
contributions to the derivative, one from the energy expression
$\langle \Psi | \hat{H} | \Psi\rangle$ and the other from the
orthogonality constraint. We will work out only  the derivatives
with respect to the left site-functions
explicitly as similar expressions hold for derivatives with respect to
the right site-functions. The derivative of the energy expression is
\begin{align}
&\partial/\partial L^{n_q l_{q-1}}_{ l_q} \langle \Psi | \hat{H} |
\Psi\rangle\nonumber\\ = &
\left(L^{n_1}_{l_1} \ldots \cancel{L^{n_q l_{q-1}}_{l_q}} 
\ldots C^{l_{p-1} n_p r_{p+1}}
  \ldots R^{n_k}_{r_{k-1}} \right) \times \nonumber \\ & H_{n_1 
 \ldots n_k}^{n_1^\prime \ldots n_k^\prime}
\left(L_{n_1^\prime}^{l_1^\prime} \ldots L_{n_q^\prime
  l_{q-1}^\prime}^{l_q^\prime} \ldots
C_{l_{p-1} n_p r_{p+1}}
\ldots R_{n^\prime_k}^{r_{k-1}} \right)
\nonumber \\ = &
W^{n_q^\prime l_{q-1}^\prime l_q }_{n_q l_{q-1} l_q^\prime}
L_{n_q^\prime l_{q-1}^\prime}^{l_q^\prime} \label{eq:energy_derivative}
\end{align}
while the derivative of the orthogonality
constraint is  
\begin{align}
-\partial/\partial \mathcal{L}^{n_q l_{q-1}}_{ l_q} \sum_{m<p}  
\mu_{l_m}^{l_m^\prime} \left( L^{l_m}_{l_m l_{m-1}} L^{n_m l_{m-1}}_{l_m^\prime} -
\delta^{l_m}_{l_m^\prime} \right)\nonumber \\ =-\mu_{l_q^\prime}^{l_q} L_{n_q
l_{q-1}}^{l_q^\prime} \label{eq:constraint_derivative}
\end{align}
and thus at the minimum, where $\partial \mathcal{L}/\partial L^{n_q l_{q-1}}_{ l_q}=0$,
\begin{align}
W^{n_q^\prime l_{q-1}^\prime l_q }_{n_q l_{q-1} l_q^\prime}
L_{n_q^\prime l_{q-1}^\prime}^{l_q^\prime} = 
\mu_{l_q^\prime}^{l_q} L_{n_q
l_{q-1}}^{l_q^\prime} \label{eq:l_cond}
\end{align}

Now the minimising condition (\ref{eq:l_cond}) does not immediately resemble solution conditions
2. and 3. for the canonical site-functions from the convergence of the
sweep algorithm. To demonstrate the equivalence,
we first recall that any minimum of the canonical Lagrangian
(\ref{eq:ortho_lagrange}) \textit{is also a minimum of the simple Lagrangian} (\ref{eq:dmrg_lagrangian})
in section \ref{sec:dmrg_lagrange} that did not have
the additional orthogonality constraints. This is because we can always insert
transformations as in (\ref{eq:wf_equivalence}) to convert a general
matrix product state (\ref{eq:mps}) to a DMRG canonical form
(\ref{eq:renorm_wf}), and such transformations do not change the
energy or wavefunction normalisation appearing in the
simple Lagrangian (\ref{eq:dmrg_lagrangian}). Thus, given some set of
$L$, $C$, $R$ that minimise the canonical Lagrangian (\ref{eq:ortho_lagrange}), these all
satisfy  site Fock equations as in  (\ref{eq:Fock}). Then, we
can substitute the Fock equation (\ref{eq:Fock}) in the energy
derivative (\ref{eq:energy_derivative}), and 
we find for  $W^{n_q l_{q-1} l_q}_{n_q^\prime
  l_{q-1}^\prime l_q^\prime}$
\begin{align}
W^{n_q l_{q-1} l_q}_{n_q^\prime
  l_{q-1}^\prime l_q^\prime}=
E 
\left(L^{n_1}_{ l_1} \ldots \cancel{L^{n_q l_{q-1}}_{l_q}} \ldots
C^{l_{p-1} n_p r_p}  \ldots R^{n_k}_{ r_k} \right) \times \nonumber \\
 \delta_{n_1 
 \ldots n_k}^{n_1^\prime \ldots n_k^\prime}
\left(L_{n_1^\prime}^{ l_1^\prime} \ldots \cancel{L_{n_q^\prime l_{q-1}^\prime}^{l_q^\prime}} \ldots 
C_{ l_{p-1}^\prime n_p^\prime r_p^\prime}  \ldots R_{n_k^\prime}^{
  r_{k-1}^\prime} \right) \label{eq:energy_deriv2} 
\end{align}
Next, we transform the wavefunctions appearing in
(\ref{eq:energy_deriv2}) through the wavefunction transformation
(\ref{eq:wave_trans}),  so that the $C$ site-function is associated with site $q+1$, i.e. 
\begin{align}
L^{n_1}_{ l_1} \ldots \cancel{L^{n_q l_{q-1}}_{l_q}} \ldots
C^{l_{p-1} n_p r_p}  \ldots R^{n_k}_{ r_k} \nonumber \\
= L^{n_1}_{ l_1} \ldots \cancel{L^{n_q l_{q-1}}_{l_q}} C^{l_q n_{q+1} r_{q+1}}
  \ldots R^{n_k}_{ r_k}  \label{eq:wf_rewrite}
\end{align}
Also we observe that the $C$ site function $C^{l_q n_{q+1} r_{q+1}}$ can always be decomposed into the product of a
matrix with an orthogonal matrix which we recognise as $R^{n_{q+1}
  r_{q+2}}_{r_{q+1}}$
\begin{align}
C^{l_q n_{q+1} r_{q+1}} = c^{l_q r_{q+1}} R^{n_{q+1}
  r_{q+2}}_{r_{q+1}} \label{eq:wf_decomp}
\end{align}
Finally at the minimum of the Lagrangian, all the $L$ and $R$
site-functions are orthogonal in the sense of (\ref{eq:ortho}) and
thus we can evaluate the contracted products of the $L$ site-functions and the
$R$ site-functions appearing in (\ref{eq:energy_deriv2}) explicitly
(substituting (\ref{eq:wf_decomp}) for $C^{l_q n_{q+1} r_{q+1}}$)
\begin{align}
\left(L^{n_1}_{l_1} \ldots L^{n_{q-1} l_{q-2}}_{l_{q-1}} \right)
\times \nonumber\\
\delta_{n_1  \ldots n_q}^{n_1^\prime \ldots n_q^\prime}
\left(L_{n_1^\prime}^{l_1^\prime} \ldots {L_{n_{q-1}
    l_{q-2}^\prime}^{l_{q-1}^\prime}} \right) = 
\delta_{l_{q-1} n_q}^{ l_{q-1}^\prime n_q^\prime}  \nonumber\\
\left({R}^{n_{q+1} r_{q+2}}_{r_{q+1} }
 \ldots R^{n_k}_{ r_{k-1}} \right) \times \nonumber \\
\delta_{n_{q+1}  \ldots n_k}^{n_{q+1}^\prime \ldots n_k^\prime}
\left({R}_{n_{q+1}^\prime r_{q+2}^\prime}^{r_{q+1}^\prime }
 \ldots R_{n_k}^{ r_{k-1}} \right) = \delta_{r_{q+1}}^{ r_{q+1}^\prime} \label{eq:eval_ortho}
\end{align}
We can now use all these simplifications (\ref{eq:wf_rewrite}), (\ref{eq:wf_decomp}),
(\ref{eq:eval_ortho}) to simplify the expression for the energy derivative $W^{n_q l_{q-1} l_q}_{n_q^\prime
  l_{q-1}^\prime l_q^\prime}$ (\ref{eq:energy_deriv2}). We find 
\begin{align}
W^{n_q l_{q-1} l_q}_{n_q^\prime
  l_{q-1}^\prime l_q^\prime}&= \delta^{n_q l_{q-1} }_{n_q^\prime l_{q-1}^\prime}
D^{l_q}_{l_q^\prime}
\end{align}
where $D^{l_q}_{l_q^\prime}$ is a density matrix built from the
coefficients $c^{l_q r_{q+1}}$
\begin{align}
D^{l_q}_{l_q^\prime} & = c^{l_q r_{q+1}} c_{l_q^\prime r_{q+1}}
\end{align}
and the Lagrangian minimising condition from the energy derivative (\ref{eq:l_cond}) becomes
\begin{align}
D^{l_q}_{l_q^\prime} L^{l_q^\prime}_{n_q l_{q-1}} =
\mu^{l_q}_{l_q^\prime} L^{l_q^{\prime
     }}_{n_q l_{q-1}} \label{eq:d_solution}
\end{align}
This minimising condition (\ref{eq:d_solution}) is in fact a density matrix solution condition very
similar to conditions 2., 3. arising from  convergence of the DMRG sweep
algorithm. To see the connection explicitly we recognise that  the density matrix $D^{l_q}_{l_q^\prime}$ (constructed from
$c^{l_q r_{q+1}}$) is related to the density matrix in the sweep
algorithm  $\Gamma^{n_q l_{q-1}
  }_{n_q^\prime l_{q-1}^\prime }$ (constructed from $C^{l_{q-1} n_q r_q}$)
in a simple way using eqn. (\ref{eq:wf_decomp})
\begin{align}
\Gamma^{n_q l_{q-1} }_{n_q^\prime l_{q-1}^\prime } = L_{l_q}^{  n_q l_{q-1}
} D^{l_q}_{l_q^\prime} L^{l_q^\prime}_{n_q^\prime l_{q-1}^\prime } \label{eq:dm_relation}
\end{align}
Next, we multiply
(\ref{eq:d_solution}) on both sides with the $L$ site function
\begin{align}
\left(L^{n_q l_{q-1}}_{l_q} D^{l_q}_{l_q^\prime}
L^{l_q^\prime}_{n_q^\prime l_{q-1}^\prime}\right) L^{n_q^\prime l_{q-1}^\prime}_{l_q^{\prime
   \prime}}
&= L^{n_q l_{q-1}}_{l_q}
\mu^{l_q}_{l_q^\prime} \left(L^{l_q^{\prime
     }}_{n_q^\prime l_{q-1}^\prime}  L^{n_q^\prime l_{q-1}^\prime}_{l_q^{\prime
  \prime}}\right)
\end{align}
Substituting in (\ref{eq:dm_relation}) for the first bracketed term and using the orthogonality of
the $L$ site-functions (\ref{eq:ortho}) for the second bracketed term this becomes
\begin{align}
E \Gamma^{n_q l_{q-1}}_{n_q^\prime l_{q-1}^\prime} L^{n_q^\prime
  l_{q-1}^\prime}_{l_q^{\prime\prime}}&  = L^{n_q
  l_{q-1}}_{l_q} \mu^{l_q}_{l_q^{\prime\prime}}
\end{align}
This is now \textit{identical} to the density matrix eigenvector condition
(\ref{eq:l_cond}) (up to a multiplicative  factor of $E$) if we simply perform a unitary transformation to
diagonalise $\mu$ such that
\begin{align}
\mu^{l_q}_{l_q^\prime} = E w^{l_q} \delta^{l_q}_{l_q^\prime}
\end{align}
Thus we have arrived at our final result: the density matrix
eigenvector condition of the sweep algorithm, which defines the $L$ and
$R$ site functions in the canonical form of the DMRG wavefunction, is
equivalent to minimising the canonical Lagrangian (\ref{eq:ortho_lagrange}), up
to a unitary transformation of each site function which does not affect the energy but
which diagonalises the matrix of Lagrange multipliers $\mu$. The DMRG
site-functions are thus ``canonical'' site functions in a way
analogous to the canonical  Hartree-Fock orbitals, which are similarly obtained from any
energy minimising set of orbitals, by performing a unitary transformation to
diagonalise the orthonormality constraints \cite{szabo1989mqc}.

\section{Conclusions}

We have shown that we can write down Lagrangians which on minimisation
yield optimal density matrix renormalisation group (DMRG)
wavefunctions in a variational sense. In particular, we have
demonstrated the equivalence between minimising a  canonical form of 
Lagrangian, and converging the DMRG energy through the original sweep
algorithm, up to certain unitary transformations of the variational
parameters in the DMRG wavefunction which leave the Lagrangian invariant. With an increasing understanding
of the DMRG from a wavefunction ansatz perspective, it is natural to 
look towards developing analytic derivative and response techniques as
are available for other kinds of wavefunction ansatz in quantum
chemistry. The results presented here are a first step in that direction.

\bibliography{dmrg-lagrange}
\end{document}